# Analyses of Statistical Structures in Economic Indices

Frank W. K. Firk

*The Henry Koerner Center for Emeritus Faculty,*

*Yale University, New Haven, CT 06520, USA*

A*bstract*:

The complex, time-dependent statistical structures observed in the Dow Jones Industrial Average on a typical trading day are modeled with Lorentzian functions. The resonant-like structures are characterized by the values of the basic ratio, $<\Delta t>/<T>$, where $<\Delta t>$ is the average lifetime of the individual states associated with a given structural form and $<T>$ is the average interval between adjacent states. Values of $<\Delta t>$ and $<T>$ are determined for three structural forms characterized by the values $<T>$ = 50 – 100 seconds (the fine structure), ≈10 minutes and ≈1 hour (the intermediate structure I and II).  During the trading day the values of the ratio $<\Delta t>/<T>$ associated with the fine structure of the index are found to lie in the narrow range from 0.49 to 0.52.  This finding is characteristic of the highly statistical nature of many-body systems typified by daily trading.  It is therefore proposed that the value of this ratio, determined in the first hour-or-so on a given day, be used to provide information concerning the likely performance of the fine, statistical component of the index for the remainder of the trading day.  The intermediate structures have values of the ratio $<\Delta t>/<T> \approx 0.6$ and therefore they can be analyzed as individual states.

*Keywords*: Analytical economics; Lorentzian analyses of statistical structures in the Dow Jones Industrial Average; basic parameters of economic indices.

## 1. INTRODUCTION

A basic parameter in systems that exhibit time-dependent resonant-like structures is the ratio (average lifetime (width) $<\Delta t>$ of the individual states)/(average interval $<T>$ between adjacent states).  If $<\Delta t>/<T> < 1$, the states are clearly separated, and the parameters associated with the individual states can be determined.  In the present analyses, *intrinsic statistical structures* observed in the daily Dow Jones Industrial Average (DJIA) on a typical trading day are analyzed using Lorentzian forms.  Each given state is characterized by its maximum value and its full width at half-height, $\Delta t$; the individual values of $\Delta t$ and T of the states are determined by fitting the observed indices.

The three analyzed structural forms are associated with long, medium, and short-range time-correlations in the complex trading options that take place among the innumerable elements that contribute to an economic index. Quantitative trading strategies require an understanding of the form of the time-dependence of the fluctuations (resonant-like states); the Lorentzian model used here provides an analytical means for studying the all-important *lifetime* of each state[1]. (Although other mathematical forms such as Gaussians or Poissonians can be used to model economic indices, they do not afford the straightforward algebraic interpretation of the *lifetime* associated with Lorentzians).  In the present analysis, the lifetime $\Delta t$ includes the formation and decay time of the state.

Analyses of the statistical structural forms of the DJIA reported on April 3, 2012 are presented.  The individual and average parameters of the structures are determined for the indices.  Conclusions that are based on the statistical properties of observed states are presented.  The results show that intrinsic, statistical properties of the states do, indeed, exist; *these properties are essentially independent of the external market forces that have unpredictable and uncontrollable effects on the indices.*

## 2. METHOD: MULTI-LEVEL LORENTZIAN ANALYSES

To model the three basic structures in the daily DJIA the following Lorentzian form is used:

$$lzn(X) = M_0 /(1 + X^2) \qquad (1)$$

where
  $X = (2/\Delta t)(t - t_0)$, a dimensionless variable,
  $t$ is the time variable,
  $t_0$ is the central time of the symmetric function,
  $M_0$ is the maximum value of the function and
  $\Delta t$ is the full width at half maximum of the function.

The area, A, under the Lorentzian is
  $$A = \pi(M_0 \Delta t)/2. \qquad (2)$$

If the average (adjacent) interval between states is $<T>$ then the average value $<lzn(X)>$ of the function $lzn(X)$ is directly related to the ratio $<\Delta t>/<T>$.

## 3. CALCULATION

### 3.1. *Preparing the data*

We are concerned with a quantitative analysis of the highly structured form of the DJIA reported on April 3, 2012; the data are shown in one-minute intervals:

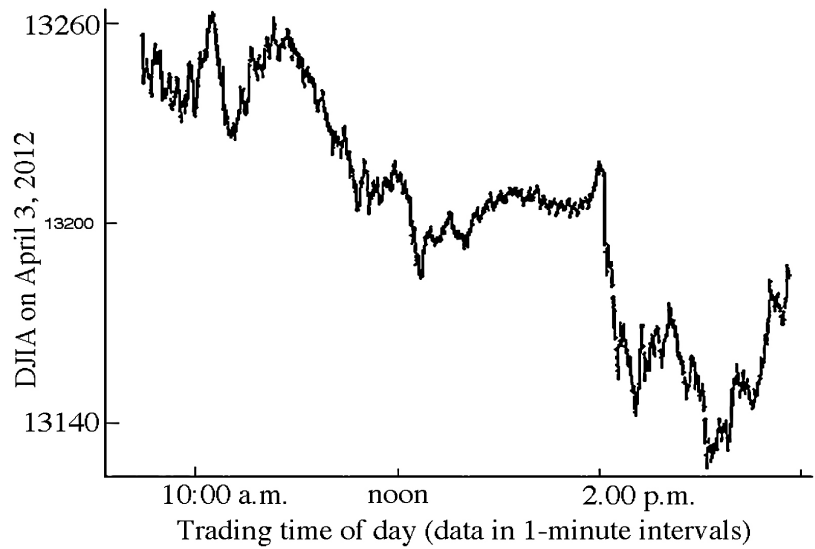

Fig.1. The 1-minute interval data of the DJIA on April 3, 2012.

The analysis requires an understanding of the basic components of the index; they are:

1. Gross structure with a characteristic time-scale of many hours,
2. Intermediate structure that appears in two distinct patterns:
    I – with an average adjacent interval of about one hour,
    II – with an average adjacent interval of about ten minutes,
3. Fine structure with a characteristic time-scale of one-to-two-minutes.

Typical intermediate components and a fine structure component are shown in Fig. 2.

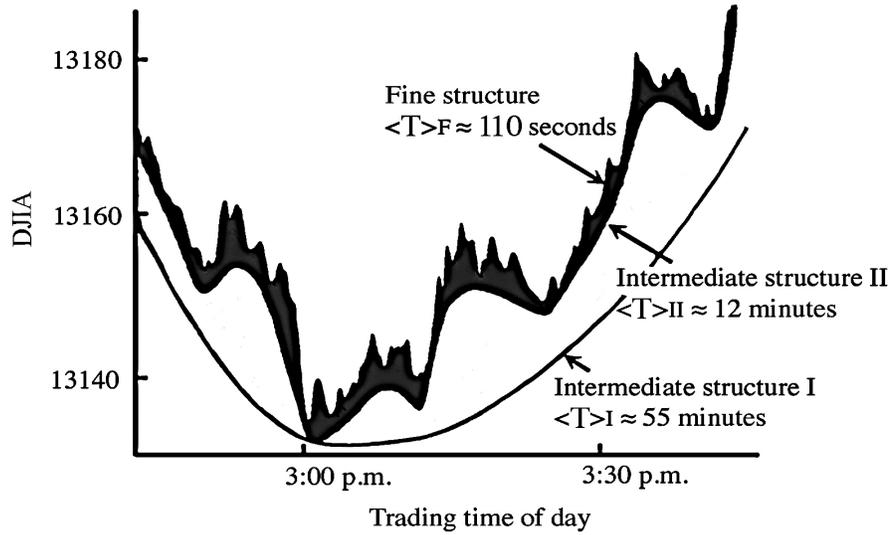

Fig. 2. The two intermediate components, and the fine structure component of the DJIA. The data are shown in 30-second intervals. The average interval of states is $<T>_I \approx 55$ min, $<T>_{II} \approx 12$ min, and $<T>_F \approx 110$ sec.

The process of normalization is continued in order to obtain the Lorentzian forms of the intermediate structures I and II, and the fine structure. To achieve adequate time-resolution, the process requires data in ten-second-intervals the individual points then can be averaged to give suitable statistical accuracy.

The fine structure data during a one-hour time period is shown in Fig. 3.

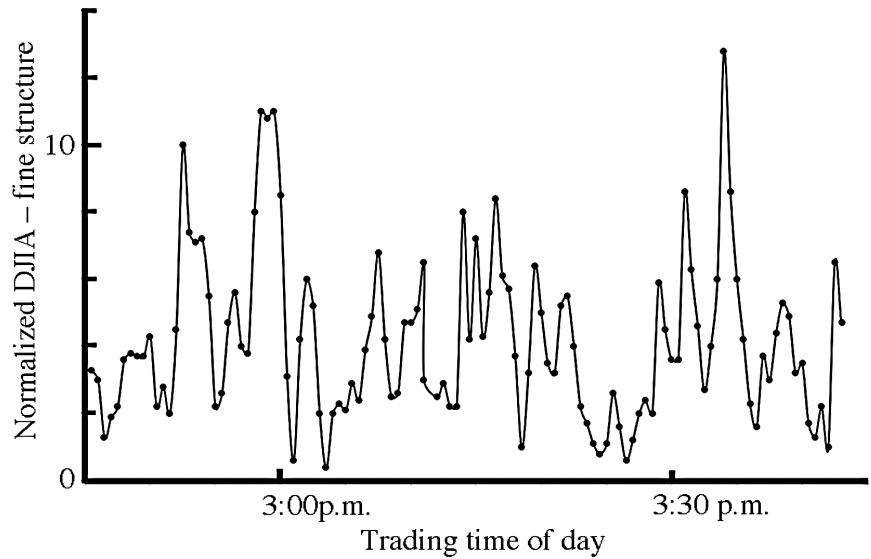

Fig. 3. The fine structure of the DJIA, shown in 30-second intervals. The average interval is $<T>_F \approx 110$ seconds.

The normalized curves of the DJIA are in forms that can be analyzed using a multi-state Lorentzian model.

## 4. RESULTS: AN ANALYSIS OF THE DJIA DATA ON APRIL 3, 2012

### 4.1. *Intermediate structure I*

A fit to the ten-minute data is shown in Fig. 4. An iterative method is used to analyze the overlapping, Lorentzian-modeled states (Eq. (1)).

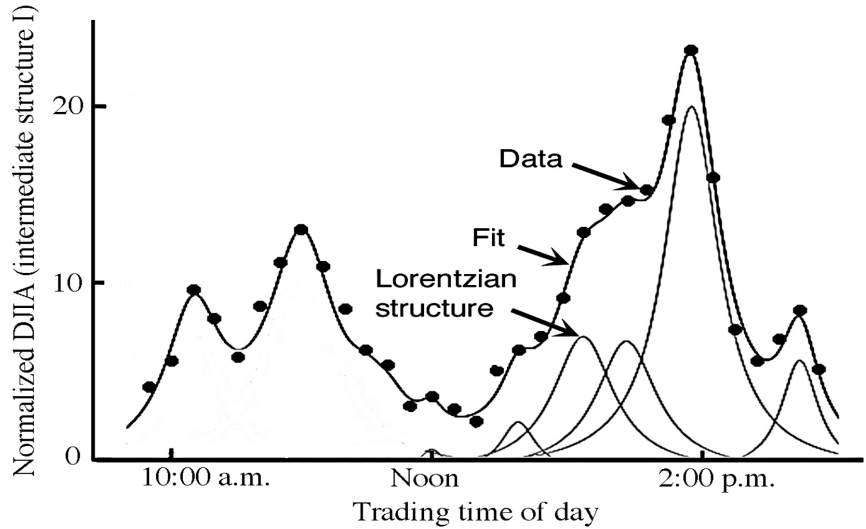

Fig. 4. Intermediate structure I: a multi-level Lorentzian analysis of the 10-minute data throughout the bulk of the trading day. The individual Lorentzians during a two-hour period are shown.

The value of $\langle \Delta t \rangle / \langle T \rangle \approx 0.6$; the relatively small number of intermediate I states during the trading day limits the accuracy of this value.

### 4.2. *Intermediate structure II*

A fit to the one-minute data in the time interval 2:45 to 3:45 p.m. is shown in Fig. 5.

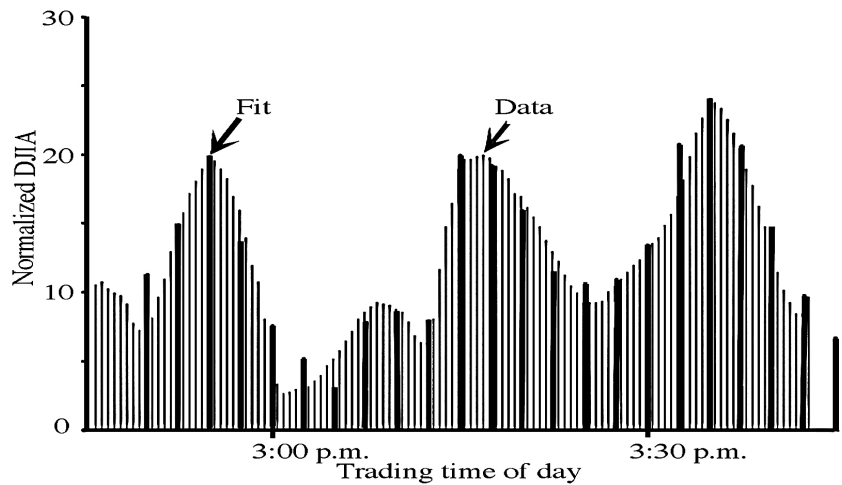

Fig. 5. A fit to the intermediate structure II during a 1-hour period. The data are shown in 30-second intervals, and the fit is shown in 150-second intervals.

The value of $<T>_{II}$ for the intermediate II structure is obtained by averaging over the entire period of the trading day; it is $<T>_{II}$ = 12 minutes; the value of $<\Delta t>/<T>_{II} \approx 0.6$.

### 4.3. *Fine structure*

The fit to the fine structure in the time interval 2:45 to 3:45 p.m. is shown in Fig. 6.

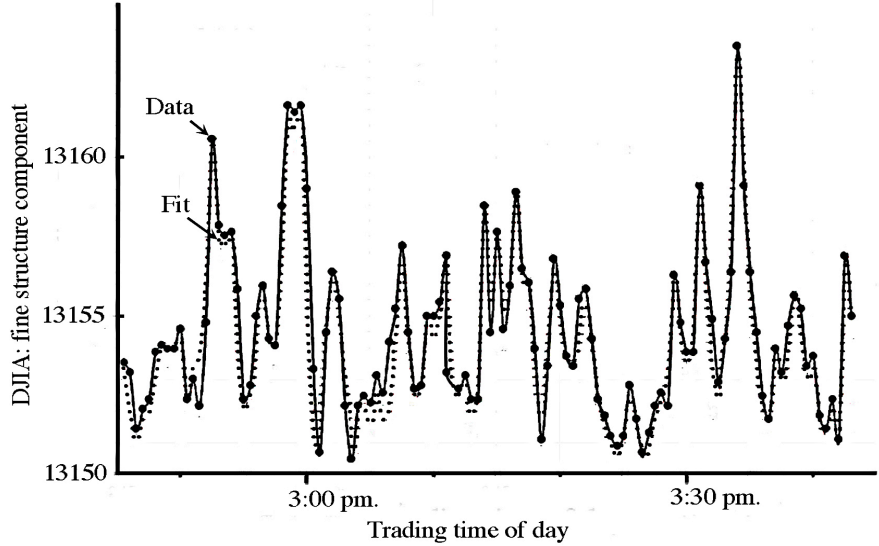

Fig. 6. The dashed line illustrates the multi-state Lorentzian fit to the fine structure data from 2:45 to 3:45 p.m.

The average value of the adjacent interval is $<T>$ = 110 seconds, and the average value of the width is $<\Delta t>$ = 57 seconds. The key ratio $<\Delta t>/<T>$ = 0.52. The analysis was repeated at four successive hourly intervals throughout the trading day and the following values of the ratio were determined: 0.49, 0.51, 0.50, 0.51, respectively.

The width and interval distributions associated with the analyzed states are shown:

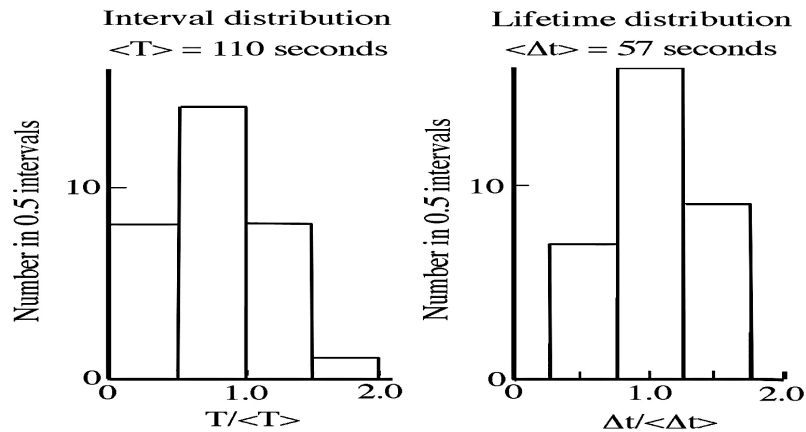

Fig.7. The interval and lifetime distributions associated with 32 fine structure states.

These distributions are consistent with the underlying statistical nature of fine structure states; they are analogous to the statistical distributions discussed in the standard work of Porter[2].

## 5. DISCUSSION: STATISTICAL BASIS OF THE ANALYSIS

### 5.1. *Statistical distributions of the states*

The *spacing distribution* of adjacent states in a many-body system like the DJIA is expected to be of a Wigner[3] form; the probability distribution function is
$$p(x) = (\pi/2) \cdot x \cdot \exp\{-\pi x^2/4\} \qquad (3)$$
where $x$ = spacing /average spacing (a dimensionless variable). This expectation is borne out by the results of Plerou et al.[4]

The *widths* of the states belong to chi-squared distributions[2]:
$$\mathrm{chi}(x; n) = \Gamma^{-1}(n/2)(n/2<x>)^{n/2} \cdot x^{(n-2)/2} \cdot \exp\{-nx/2<x>\} \qquad (4)$$
where $\Gamma$ is the incomplete gamma function.

If n = 2, the Porter-Thomas[2] distribution is obtained (for single channel processes); it is an exponential. For larger values of n, the distribution goes to zero as $x$ goes to zero: it is a more narrowly defined function, typical of many-channel decay processes. This is the case for the data shown in Figure 7.

### 5.2. *Statistical distributions of the fine structure states*

The arguments used to justify the Wigner spacing distribution of adjacent states in nuclear resonance reactions can, with suitably modified terminology, be used to justify the Wigner spacing distribution of the present fine structure states. The limited data set considered in Section 3.4 is consistent with such a distribution. Previous work (Plerou et al.[4]) that covered an economic index over a period of years found good agreement with a Wigner form for the longer-term interval distribution. Effects of long-term correlations on the form of the distribution also were found.

The statistical nature of the spacings of the fine structure states is a consequence of the innumerable, uncorrelated trading options associated with the elements that contribute to an economic index. The narrow range of lifetimes of the fine structure states is the result of the inherent decision-making times of traders, and of the rate of flow of information available to them.

## 6. CONCLUSIONS

Analyses of the DJIA on April 3, 2012 have been carried out using multi-level Lorentzian functions to model the index. In order to analyze the data it is necessary to understand, and to quantify, the basic structural components of the index. The structures that are necessary to represent a total index are 1) gross structures with time scales of many hours; 2) intermediate structures that appear in two distinct forms: i) structures with time scales of about one hour, and ii) structures with time scales of about ten minutes and 3) fine structures with time scales in the range 30-to-150 seconds.

The fine structure states are excellent examples of discrete states with spacing and lifetime distributions that are well understood in the theory of statistical fluctuation phenomena (Porter[2]). The parameters of the defining Lorentzians are used to obtain the ratio $<\Delta t>/<T>$. The fine structure states studied have typical ratios of 0.49, 0.51, 0.50, 0.51 and 0.52 in five hourly intervals throughout the day. The observed near constancy of the ratio is evidence for the purely statistical nature of the fine structure states – the value is largely unaffected by the uncontrollable, time-dependent market forces that can have such dramatic effects on the value of an economic index when studied over periods of many hours or days.

The fine structure observed can have an average interval as small as one minute, and an average width of thirty seconds. The value of the fine structure ratio depends upon the average load, bandwidth, processing capacity, and delays in buy and sell orders (the latency) associated with the fine structure on a given day.

Within the statistics imposed by the limited number of states in the present study the intermediate-I and -II structures have values $\langle\Delta t\rangle/\langle T\rangle \approx 0.6$; they are states that also can be modeled using well-defined Lorentzians. More extensive analyses are required to study in detail their statistical properties in order to uncover the possible effects of local correlations and external market forces on their structural forms.

It is proposed that the values of $\langle\Delta t\rangle/\langle T\rangle$ for the fine structure of the states, on a given day, be obtained by analyzing the data in the first one- to two-hours of daily trading, and that these values be used as *predictors* of their values for the remaining trading time of the day. This procedure assumes that there is sufficient statistical inertia in the fine structure associated with trades on the given day; the present analysis provides support for this assumption. Knowledge of the *lifetimes* and *adjacent intervals* of the states that make up the basic components of an economic index is important in developing any quantitative investment strategy. The analysis presented here can be applied to other economic indicators (daily currency exchange rates, for example) in which fine structure of a statistical nature is observed.